\documentclass[final,1p,times]{elsarticle} 
\usepackage{graphicx} 
\usepackage{amssymb} 
\usepackage{amsthm} 

\journal{Nuclear Physics A} 
\begin{document} 

\begin{frontmatter} 


\title{Measurement of $\pi^{0}$ and $\gamma$ in d+Au collisions at $\sqrt{s_{NN}}=$\,200\,GeV by PHENIX experiment}

\author{Ond\v{r}ej Chv\'{a}la$^{a}$ for the PHENIX collaboration}

\address{$^{a}$ University of California -- Riverside, 900 University Ave, Riverside CA 92521, USA}

\begin{abstract} 
Previous results 
indicated that high $p_T$ particle suppression in Au+Au interactions is a final state effect, since R$_{dA}$ ratios were compatible with unity, albeit within large experimental errors.
It is important to test this conclusion to higher precision since the modification of structure functions may be involved.
Recent d+Au data taken in 2008 improve the integrated luminosity by about a factor of thirty compared to the 2003 data.
A more precise measurement of both $\pi^0$ and $\gamma$ at higher $p_T$ will shed new light on whether the initial state in the heavy nuclei is modified.


\end{abstract} 

\end{frontmatter} 


\section{Introduction}\label{intro}
The PHENIX experiment at RHIC is designed for high rate measurement 
of electromagnetic probes, specifically direct photons $\gamma$ 
and neutral pions $\pi^{0}$ at mid-rapidity ($|\eta|<$\,0.35). 
The high event rate allows unique access to rare probes, such as particles produced at very high transverse momentum ($p_T$).

Early results for the nuclear modification factors R$_{AA}$, 
the spectrum in Au+Au collisions over the spectrum in p+p interactions
scaled by the respective nucleon overlap integrals, 
showed suppression of $\pi^{0}$ and charged hadron production at high-$p_{T}$ 
($p_{T} \gtrsim$\,6\,GeV/c) \cite{Adcox:2001jp,Adcox:2002pe,Adler:2003qi}.
This result, combined with the apparent lack of suppression
for high-$p_T$ direct $\gamma$ production \cite{Adler:2005ig}
has been interpreted as an indication of the formation of a dense strongly interacting medium, the sQGP, in heavy ion collisions.

Furthermore, the preliminary analysis of the high statistics 2004 Au+Au run showed a possible decrease of direct photon R$_{AA}$ above 14\,GeV/c of transverse momentum \cite{Reygers:2008pq}. Possible explanations for this observation include the isospin effect (the difference of partonic content between protons and neutrons), the EMC effect (modifications of the distribution of partons in the heavy nuclei), and the suppression of the direct photons originating from fragmenting partons which are quenched by the medium.

RHIC measurements of particle production in d+Au collisions from 2003 show no suppression of produced particles within large experimental error bars \cite{Adler:2003ii,Peressounko:2006qs}, indicating little or no modification of the initial state in gold nuclei. 
This result confirmed the attribution of the large suppression observed in central Au+Au interactions to final state effects in the sQGP medium.  

The final analysis of the 2003 data set revealed some suppression of $\pi^{0}$ production at the highest $p_T$ in the most central d+Au interactions \cite{Adler:2006wg}, however the experimental uncertainties are too large 
to estimate cold nuclear matter effects quantitatively \cite{Barnafoldi:2008ec,Zhang:2008ek}.

\section{RHIC year 2008 data set}\label{2008}
\begin{figure}[ht]
\hspace{-0.6cm}\includegraphics[width=1.05\textwidth]{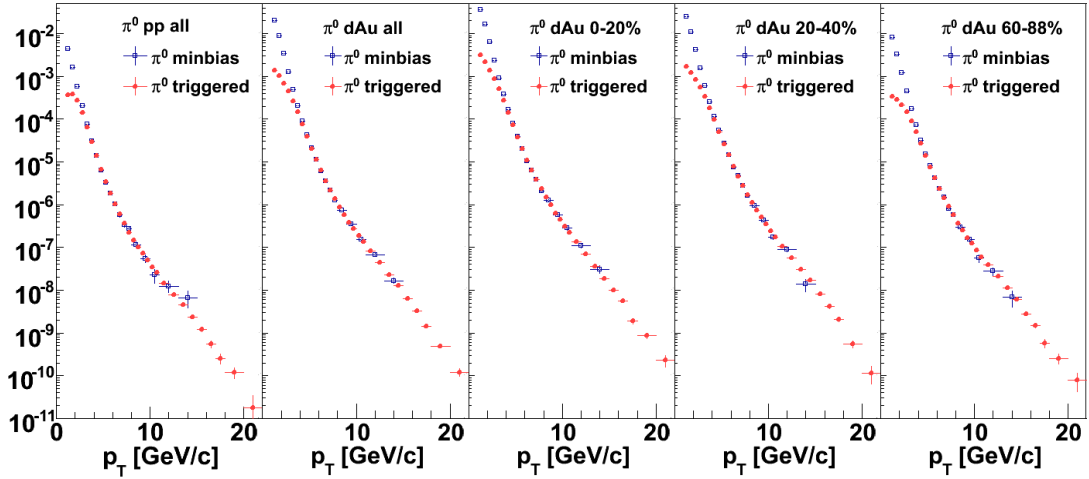}\vspace{-0.2cm}
\caption[] {Raw mid-rapidity yields per event of $\pi^{0}$ production from the 2008 data sets: 
p+p minbias, d+Au minbias, and three centrality bins in d+Au collisions using PHENIX PbSc EMCal. 
The spectra from ERT triggered runs (red) are scaled to match MB yields (blue) in 6--13\, GeV/c region of $p_{T}$.
Only statistical errors are shown.}
\label{fig:pi0}
\end{figure}

The RHIC run in year 2008 improved the total integrated luminosity of the d+Au sample 
by a factor of $\approx$30 compared to the 2003 run:
1.65$\times$10$^9$ minimum bias (MB) events, 
and 3.68$\times$10$^9$ events from a high-$p_{T}$ photon (ERT) trigger were recorded.
There were also 0.53$\times$10$^9$ MB triggered and 1.17$\times$10$^9$ ERT triggered 
p+p collisions taken in 2008. 
The uncorrected $\pi^{0}$ yields are shown in Fig.~\ref{fig:pi0}.

The improvement in total d+Au integrated luminosity can be appreciated from the relative statistical errors in the 
$\pi^{0}$ measurement using PHENIX PbSc EMCal. 
The error at $p_{T}=$\,15\,GeV/c is 32\,\% using the 2003 sample, and is reduced to 3.5\,\% in the recent high statistics run.

\vspace{-0.4cm}
\section{Conclusions}\label{conclusions}

Previous results from data taken in year 2003 indicate that the high $p_{T}$ suppression 
observed in Au+Au interactions is a final state effect since the R$_{dA}$ ratios are consistent with unity, albeit within large experimental errors. 
It is important to test this conclusion to higher precision. 
Recent d+Au data taken in 2008 improve the integrated luminosity by about a factor of thirty compared to the 2003 run.
The 2008 PHENIX measurement will shed more light on the origin of cold nuclear matter effects.

\vspace{-0.4cm}


\end{document}